\documentclass[12pt]{iopart}
\usepackage{amsfonts}
\usepackage{amssymb}

\usepackage{mathrsfs}
\usepackage{bm}
\usepackage{graphicx}
\begin{document}
\title{Matter-wave diffraction in time with a linear potential}
\author{A. del Campo and J. G. Muga}
\eads{\mailto{qfbdeeca@ehu.es},\mailto{jg.muga@ehu.es}}
\address{Departamento de Qu\'\i mica-F\'\i sica,
Universidad del Pa\'\i s Vasco, Apdo. 644, Bilbao, Spain}%
\def\la{\langle}
\def\ra{\rangle}
\def\om{\omega}
\def\Om{\Omega}
\def\vep{\varepsilon}
\def\wh{\widehat}
\def\tr{\rm{Tr}}
\def\da{\dagger}
\newcommand{\beq}{\begin{equation}}
\newcommand{\eeq}{\end{equation}}
\newcommand{\beqa}{\begin{eqnarray}}
\newcommand{\eeqa}{\end{eqnarray}}
\newcommand{\intf}{\int_{-\infty}^\infty}
\newcommand{\into}{\int_0^\infty}
\begin{abstract}

Diffraction in time of matter waves incident on a shutter
which is removed at time $t=0$  
is studied in the presence of a linear potential.  
The solution is also discussed in phase space in terms of the Wigner function. 
An alternative configuration relevant to current experiments 
where particles are released from a hard wall 
trap is also analyzed for single-particle states and for 
a Tonks-Girardeau gas.

\end{abstract}

%
%
%
%
\section{Introduction}
Diffraction in time (DIT) was studied first by Moshinsky 
\cite{Moshinsky52,Moshinsky76}, and a flurry of both experimental 
and theoretical work 
investigating similar transient effects and setups 
have been carried out ever since. 
(For a brief, recent review see \cite{DM05}.) 
Quantum 
temporal oscillations of matter waves released from a shutter 
or confinement region constitute the hallmark of the effect. 

Remarkably, all the theoretical studies have dealt with 
the quantum dynamics in free space 
or potentials with finite support \cite{Kleber94,GR97,GV01}.
However, the presence of a linear potential is ineludible for some experimental 
setups. This is actually the case for the first  experimental 
observation of DIT, which made explicit use of gravity 
on a system of ultra cold atoms \cite{SSDD95,ASDS96}.
Gravitational effects are also crucial in cold atom fountains 
as those used for frequency standards \cite{Salomon}.
In addition, a linear potential arises from the interaction 
of a charged particle with an electric field, but  
it is within the field of atom optics that linear potentials, 
generally created by a magnetic field,
have  received a great deal of attention for ultracold atom manipulation, 
as a plugging potential for loading schemes, 
or aimed to cloud focusing, cooling, and other key operations \cite{FKS02,MS99}.

One of the natural configurations to deal with such situations in 
the stationary regime, namely that of a point-like and 
Gaussian sources, has already been discussed in excellent agreement with 
photo-detachment and atom-laser experiment \cite{Kramer03}.
In this work we generalize the Moshinsky shutter problem, i.e., a cut-off 
plane wave released at time $t=0$, by turning on a linear potential 
at that instant. 
The exact solution is derived in section \ref{ditlinear} together with its 
Wigner function. 
A related configuration, namely, that of particles initially trapped in 
a rectangular box is analyzed in section \ref{hwt}.
The experimental build up of 
all-optical hard-wall traps \cite{MSHCR05} has excited a great deal of attention for such 
geometries both at single \cite{GK76, Godoy02} and many-particle level 
\cite{Gaudin71,Cazalilla02, BGOL05,DM05b}.
In particular, experiments are in view to study the quantum dynamics 
in the Tonks-Girardeau regime with Fock states of low particle number, $N$ \cite{Mark}.
The experimental observation of DIT with a Bose-Einstein 
condensate \cite{CMPL05} 
is in excellent agreement with the dynamics entailed in the time-dependent 
Schr{\"o}dinger equation, showing that the condensate evolves as a free 
non-interacting particle gas after few millisecond of expansion.
Indeed, the importance of the non-linearity during free expansion has been studied 
in full detail for the case of a Tonks-Girardeau gas in \cite{DM05b} and shown 
to dominate in a time scale shorter than $t_{N}=mL^{2}/(2N\pi\hbar)$, 
$L$ being the size of the initial trap and $m$ the mass of the particle.
The expansion of a Tonks-Girardeau gas will be analyzed in section \ref{tg}.

%

\section{Diffraction in time in free space}\label{ditfree}

The paradigmatic setup for DIT corresponds to a beam  
of particles incident on a shutter represented by an infinite potential barrier.
The relevance on the diffraction pattern of the reflectivity of the barrier 
\cite{Moshinsky52} as well as different types of beam have been
thoroughly investigated in \cite{DM05}.
For simplicity, we will assume here a totally absorbing barrier so that 
the initial state is given by a cut-off plane wave,
\beq 
\label{eq1}
\psi(x,t=0)=e^{i p x/\hbar}\Theta(-x). 
\eeq
The solution to the time dependent Schr{\"o}dinger equation 
\cite{Moshinsky52, Moshinsky76} reads
\beq
\psi(x,t)=M(x,p/\hbar,\hbar t/m):=\frac{e^{i\frac{m x^{2}}{2 t\hbar}}}{2}w(-z),
\eeq
with 
\beq
z=\frac{1+i}{2}\sqrt{\frac{t}{m\hbar}}\left(p-\frac{mx}{t}\right),
\eeq
and 
the so called Faddeyeva function 
\cite{Faddeyeva,Abram} $w$ is defined as
\beq
w(z):= e^{-z^{2}}{\rm{erfc}}(-i z)=\frac{1}{i\pi}\int_{\Gamma_{-}}du
\frac{e^{-u^{2}}}{u-z},
\eeq
where $\Gamma_{-}$ is a contour in the complex $z$-plane which goes from
$-\infty$ to $\infty$ passing below the pole.
After \cite{Moshinsky51, Moshinsky52}, $M(x,k,t)$ has been named 
the Moshinsky function.
The hallmark feature of this phenomenon is that, at variance with the 
classical solution, the probability density 
presents genuine quantum oscillations in time and space.
Notice that a point of constant probability $P$
propagates as
\beq
x_{P}(t)=\frac{p t}{m}-\sqrt{\frac{\pi\hbar t}{m}} u_{0, P},
\eeq
$u_{0, P}$ being a given constant.
This result will be compared with the case where the evolution 
takes place in the presence of a linear potential.

\section{Diffraction in time with a linear potential}\label{ditlinear}
We next study the DIT in the presence  of a linear potential. 
Consider a beam of particles incident from the left as before on
a totally absorbing shutter located at position $x=0$, see (\ref{eq1}). Suddenly,
the shutter is removed at time equal zero, and a linear potential is switched on. 
Such model can simulate the quantum dynamics of a confined charged particle, 
with a homogeneous electric field turned on at the time the infinite potential 
wall at the origin is removed. Also, it can be applied to a particle in a 
gravitational 
field; in that case the cut-off plane wave may be used as a basis element 
to describe the actual initial state.

The evolved wave function obeys the integral equation
\beq\label{inteq}
\psi(x,t)=\int_{-\infty}^{\infty}dx' K(x,t\vert x',0)\psi(x',0), 
\eeq
where 
$K(x,t\vert x',t')$ is the propagator for a Hamiltonian with a general 
linear potential of the form 
\beq\label{hamilt}
\hat{H}=\frac{p^{2}}{2m}+f x,
\eeq
with $f$ constant. It is indeed well known \cite{GS98}, 
\beq\label{propgrav}
K(x,t\vert x',t')=\sqrt{\frac{m}{2\pi i\hbar (t-t')}}\,
e^{i\frac{m(x-x')^{2}}{2\hbar (t-t')}
-i\frac{f(x+x')(t-t')}{2\hbar}-i\frac{f^{2} (t-t')^{3}}{24\hbar}}, 
\eeq
and related to the action of the 
classical path.

\begin{figure}
\begin{center}
\includegraphics[height=5cm,angle=0]{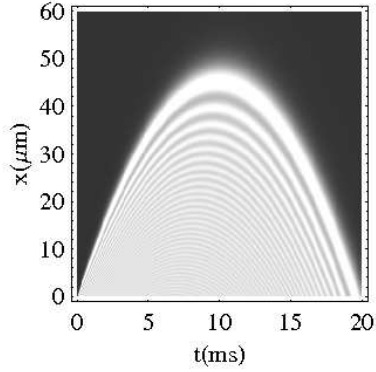}
\end{center}
\caption{\label{dit} 
Probability density on the $t-x$ plane, which clearly shows the diffraction 
in time and space for an incident beam of Rubidium atoms in a fountain configuration 
($p/m=1$ cm/s, $f/m=0.1*g$). 
The grey scale changes from light to dark as the function values decrease. 
The same color coding is used in all density plots.}
\end{figure}

Inserting (\ref{propgrav}) into (\ref{inteq}) we find 
\beq
\label{equ}
\psi(x,t)=\sqrt{\frac{m}{2\pi i\hbar t}}\,e^{-i\frac{f^{2} t^{3}}{24\hbar}
-i\frac{f t x}{2\hbar}}\int_{-\infty}^{0}dx'e^{i\frac{m(x-x')^{2}}{2\hbar t}
+i(p-ft/2)x'/\hbar}.
\eeq
It is convenient to introduce the auxiliary variables
\beq
\fl u\equiv \sqrt{\frac{m}{\pi t \hbar}}(x'-x)+\sqrt{\frac{t}{\pi m\hbar}}
(p-ft/2),\\
u_{0}\equiv \sqrt{\frac{t}{\pi m\hbar}}(p-ft/2)-\sqrt{\frac{m}{\pi t \hbar}}x, 
\eeq
to write (\ref{equ}) as 
\beq
\label{psiu}
\psi(x,t)=\frac{1}{\sqrt{2i}}\,e^{-i\frac{f^{2} t^{3}}{24\hbar}
-i\frac{f t x}{2\hbar}}\int_{-\infty}^{u_{0}}due^{i\frac{\pi u^{2}}{2}},
\eeq
which can be expressed  in terms of the Fresnel sine and cosine integrals,  
\beq
\int_{-\infty}^{u_{0}}du\, e^{i\frac{\pi u^{2}}{2}}=\frac{1+i}{2}
+\int_{0}^{u_{0}}du\,e^{i\frac{\pi u^{2}}{2}}=\frac{1+i}{2}+C(u_{0})+iS(u_{0}).
\eeq
From the last equation it is possible to map the probability density to the Cornu
spiral \cite{Moshinsky52}. 
Additionally, the following relation between the Fresnel integrals and the 
$w$-function holds \cite{Abram}, 
\beq
C(u_{0})+iS(u_{0})=\frac{1+i}{2}\left(1-e^{i\frac{\pi u_{0}^{2}}{2}}
w\left[\frac{\sqrt{\pi}}{2}(1+i)u_{0}\right]\right). 
\eeq
Then, one can make use of the Faddeyeva's identity, which follows from
the Cauchy theorem,
\beq
w(z)+w(-z)=2e^{-z^{2}},
\eeq
to end up with the result
\beqa\label{sol}
\psi(x,t) & = & \frac{1}{2}e^{-i\frac{f^{2} t^{3}}{24\hbar}
-i\frac{f t x}{2\hbar}
+i\frac{m x^{2}}{2 m \hbar}}w(-z),
\eeqa
where now  
\beq
z\equiv\frac{1+i}{2}\sqrt{\frac{t}{m\hbar}}\left(p-\frac{ft}{2}
-\frac{mx}{t}\right).
\eeq
In Fig. \ref{dit} a contour diagram of the probability density on the plane $t-x$  is plotted for a ``fountain configuration'', namely, for the case in which the 
initial velocity goes in the direction of increasing potential. 
In such representation it is particularly easy to appreciate the diffraction  
in both time and space domains. 

%
%
\begin{figure}
\begin{center}
\includegraphics[height=5cm,angle=-90]{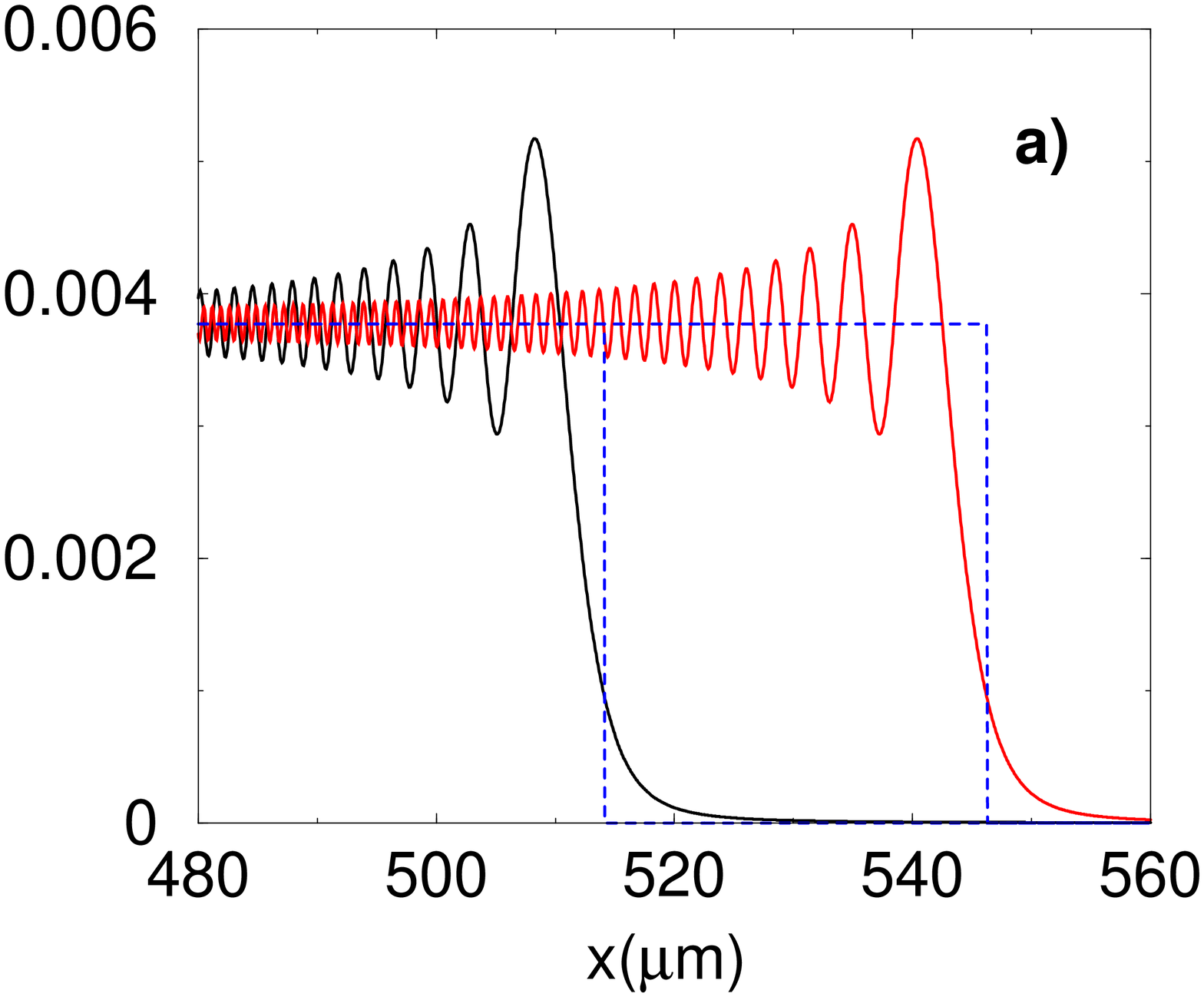}\hspace{2cm}
\includegraphics[height=5cm,angle=-90]{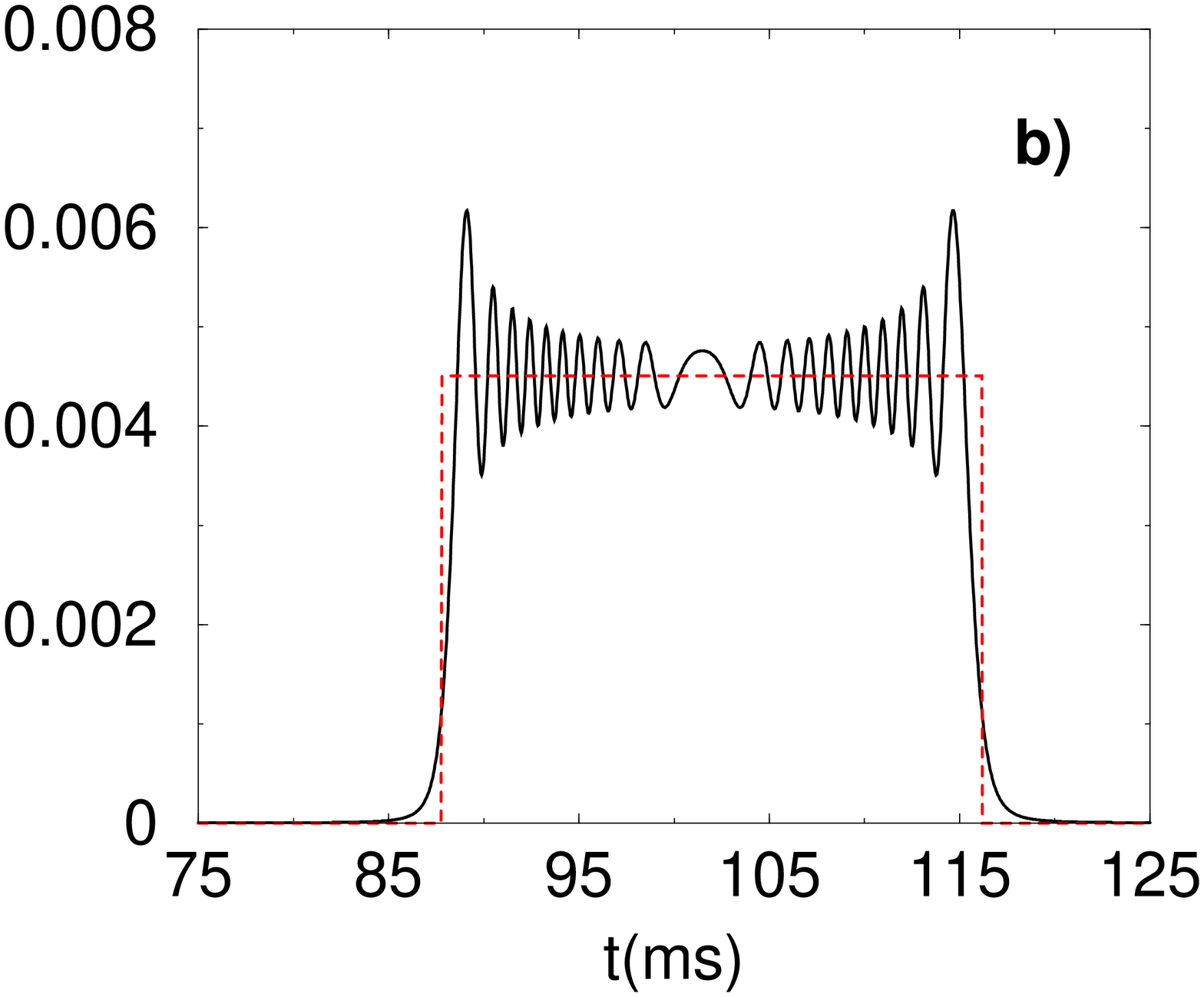}
\end{center}
\caption{\label{ditt} 
a) Probability density for both free an accelerated beam at $x>0$ with 
$p/m=5$ cm/s, $f/m=-0.1$ cm/s$^{2}$ 
(solid lines) and corresponding the classical density (dashed lines) 
$10$ ms after removing the shutter ($m=m_{Rb}$ in all the calculations). \\
b) Diffraction in time for an incident beam with $p/m=10$ cm/s 
registered by a detector located at $5$ mm from the shutter, 
$f/m=0.05$ cm/s$^{2}$. 
The dashed line corresponds to the classical density.}
\end{figure}
%
To learn about the main features of this solution, 
we compare it in Fig.\ref{ditt} with the free case.
Indeed, it is amusing to rewrite (\ref{sol}) in terms of the Moshinsky function as
\beq
\psi(x,t)= e^{-i\frac{f^{2} t^{3}}{6 m\hbar}+i\frac{ftx}{\hbar}}\,
M\left(x-\frac{f t^{2}}{2m},\frac{p}{\hbar},\frac{\hbar t}{m}\right), 
\eeq
which exactly reduces to the result of section \ref{ditfree} for $f=0$.
Notice that the dynamics is modified with respect to the free particle 
case by correcting both momentum 
and position with the classically expected time-dependent canonical 
transformation, as shown in Fig. \ref{ditt}a, up to an additional phase. 
In fact, the main difference is that, for a given $u$ the equation of motion 
for the associated point of constant probability,
and in particular for the maximum and minimum, can be deduced to be
\beqa
x_{max}(t)&=&\frac{p t}{m}-\frac{f t^{2}}{2 m}
-\sqrt{\frac{\pi\hbar t}{m}} u_{0, max},
\\
x_{min}(t)&=&\frac{p t}{m}-\frac{f t^{2}}{2 m}
-\sqrt{\frac{\pi\hbar t}{m}}u_{0, min},
\eeqa
where $u_{max}=1.2172$ and $u_{min}=1.8725$ are universal for the 
cut-off plane wave case considered here. From these equations the turning point 
in Fig. \ref{ditt} can be inferred.

As stated above, the telltale sign of DIT 
is a set of oscillations, observed after removing the shutter, 
in the probability density and which are genuinely quantum in nature. 
For diffraction in time with a linear potential, the fringe visibility 
\beq
\mathcal{V}(T)=\frac{P_{1^{st} max}-P_{1^{st} min}}{P_{1^{st} max}
+P_{1^{st} min}}
\eeq
turns out to be independent of time, for the quasi-monochromatic case, 
as in the absence of the external field. 

As pointed out before the diffraction in time admits also a representation of 
the Cornu spiral.
The classical probability density, given by a step function, intersects 
the quantum one at two different points. 
Therefore, an estimate of the width for the largest fringe can be carried 
out by considering the intersection with the classical probability density. 
The same result than without linear potential  holds, 
\beq\label{fringe}
\Delta x\simeq 0.85\sqrt{\frac{\pi\hbar t}{m}}
\eeq
\cite{Moshinsky52,Moshinsky76}
provided that the probability is exclusively $u$-dependent,
as should be clear from (\ref{psiu}), $P_{max}=1.370;
 P_{min}=0.778$.
An alternative representation which explicitly exhibits 
the diffraction pattern in time domain is shown in Fig. \ref{ditt}b, 
where the position of the detector is fixed and the signal of the 
incident beam is recorded as a function of time.
One should notice the similarity with the profile in coordinate 
representation obtained by releasing a particle from a hard-wall 
trap with $f=0$. 
This  configuration was identified as the analogue of 
diffraction in time from a slit in free space \cite{Godoy02}.
Nevertheless, due to the dispersion entailed in (\ref{fringe}) the  
pulse registered in the time domain is not symmetric.


\subsection{Wigner transform}
The Wigner function \cite{Wigner32}, 
\beq
\label{wigner}
\mathcal{W}(x,p) := \frac{1}{\pi\hbar}\int_{-\infty}^{\infty}dy\, 
\psi(x+y)^{*}\psi(x-y)e^{2ipy/\hbar}, 
\eeq
is the best known quasi joint probability distribution for position and momentum.
Nowadays, it is possible to study it experimentally as has been 
demonstrated in a series of works \cite{wigexp}.
In the context of the Moshinsky shutter,  its time evolution was derived for the 
free case problem in \cite{MMS99}.
From that result at $t=0^{+}$ or the definition (\ref{wigner}), 
it follows that for the cut-off plane wave initial condition, 
the Wigner function reads 
\beq
\mathcal{W}(x,p;p_{0},t=0^{+})=\frac{1}{\pi}
\frac{\sin[-2x(p_{0}-p)/\hbar]}{p_{0}-p}\Theta(-x),
\eeq
which clearly assumes negative values. For potentials of degree $n=2$ at most,  
the time evolved Wigner function follows the classical trajectories, 
\beq
\fl\mathcal{W}(x,p;t)=\int\!\!\!\int dx_{0}dp_{0}\delta[x-x_{cl}(x_{0},p_{0},t)]
\delta[p-p_{cl}(x_{0},p_{0},t)]\mathcal{W}(x,p;t=0).
\eeq
Alternatively one can cope with the more cumbersome calculation starting from the definition 
as in \cite{MMS99}.
For (\ref{sol}) the result is
\beqa
\fl\mathcal{W}(x,p;p_{0},t)= \frac{\sin\bigg\{\frac{2}{\hbar}\left(\frac{p t}{m}
+\frac{f t^{2}}{2m}-x\right)
(p_{0}-f t-p)\bigg\}}
{\pi(p_{0}-f t-p)}\Theta\left(\frac{p t}{m}+\frac{f t^{2}}{2m}-x\right) .
\eeqa  
%
\begin{figure}
\begin{center}
\includegraphics[height=6cm,angle=0]{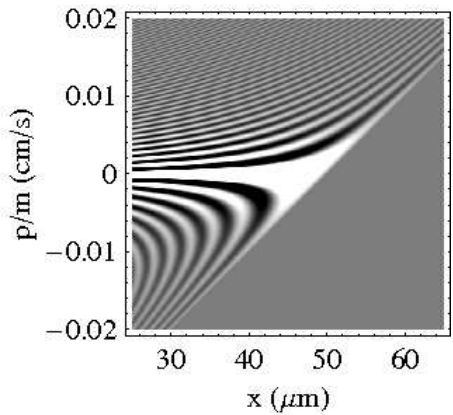}
\end{center}
\caption{\label{wignerplot} 
Wigner function for the atom fountain configuration at the classical turning point 
($p_0/m=1$ cm/s, $f/m=100$ cm/s$^{2}$, $t=10$ ms). 
The grey triangle in the right lower corner is the classically unaccessible 
region for which $\mathcal{W}(x,p;t)=0$.
}
\end{figure}
%
Fig. \ref{wignerplot} shows the Wigner function calculated for an incident 
beam exactly at the classical turning point, being thus centered around $p=0$.
Note that thanks to the step function, it is limited to the classically 
accessible region of the phase space.
Indeed, following \cite{MMS99} we can ask for the classical limit taking 
$\hbar\rightarrow 0$ to find
\beq
\mathcal{W}_{cl}(x,p;p_{0},t)  
 = \delta(p_{0}-f t-p)\Theta\left(\frac{p_{0} t}{m}-\frac{f t^{2}}{2m}-x\right).
\eeq
As expected, the correct classical equation of motion 
${x=x_{cl}(t),p=p_{cl}(t)}$ is recovered and no hint 
about the quantum oscillations -hallmark feature of DIT- remains.

\section{The hard-wall trap}\label{hwt}

In this section we consider the dynamics of a particle initially confined 
in a box. At time zero its walls are suddenly removed and a linear potential 
is switched on.  
The paradigmatic particle in a box (PIAB) has been recently 
implemented in an all optical fashion  by Raizen {\it et al.} 
\cite{MSHCR05} with a Bose-Einstein condensate
and experiments 
are in view with Fock states of low number of particles, $N<10$ \cite{Mark}.
From the theoretical point of view, the free evolution of the wavefunction 
which results after shutting off the walls was discussed long time ago within
the context of ultracold neutrons interferometry \cite{GK76}.
Later, it was reformulated by Godoy \cite{Godoy02}, who pointed out 
the analogy with the Fraunhoffer diffraction in the case of small box 
(compared to the De Broglie length), and Fresnel diffraction, 
for larger confinements. Quite recently , a generalization to an 
interacting many-body system was investigated in 
the regime of a Tonks-Girardeau gas \cite{DM05b}.

If $\chi_{[0,L]}(z)$ is the characteristic function in the interval $[0,L]$,
let us now consider the initial condition given by
an arbitrary excited state $|\psi_{n}\ra$, 
\beq
\psi_{n}(x,t)=\sqrt{\frac{2}{L}}\sin\left(\frac{n\pi x}{L}\right)
\chi_{[0,L]}(x). 
\eeq
Using the decomposition of the characteristic function
$\chi_{[0,L]}(x)=\Theta(L-x)-\Theta(-x)$, the $\psi_{n}(x,t)$ term can 
be worked out to be
\beqa
\fl\psi_{n}(x,t)&=&  
\sqrt{\frac{m}{2\pi i\hbar t}}e^{-i\frac{f t x}{2\hbar}
+i\frac{f t^{2}}{24\hbar}}
\nonumber
\\
&\times&\sum_{\alpha=\pm}\alpha\left[\int_{-\infty}^{L}dx'\,
e^{i\frac{m(x-x')^{2}}{2\hbar t}
+i\frac{p_{\alpha} x'}{2\hbar}}-\int_{-\infty}^{0}dx'e^{i\frac{m(x-x')^{2}}
{2\hbar t}+i\frac{p_{\alpha} x'}{\hbar}}\right],
\eeqa
where we have defined
\beq
p_{\alpha}\equiv\frac{\alpha\hbar n\pi}{L}-\frac{f t}{2}.
\eeq
Carrying similar steps to the ones described for the previous section, one finally 
ends up with
\beq\label{bgrav}
\fl\psi_{n}(x,t)  = \frac{1}{4 i}\sqrt{\frac{2 }{L}}\,
e^{-i\frac{f t x}{2\hbar}
+i\frac{f t^{2}}{24\hbar}}\sum_{\alpha=\pm}\alpha
\left[e^{i\frac{p_{\alpha}L}{\hbar}
+i\frac{m(x-L)^{2}}{2\hbar t}}\,w(-u_{\alpha L})
-e^{i\frac{m x^{2}}{2\hbar t}}\,\,w(-u_{\alpha 0})\right],
\eeq
where
\beq
u_{\alpha L}=\frac{1+i}{2}\sqrt{\frac{t}{m\hbar}}
\left(p_{\alpha}-\frac{m(x-L)}{t}\right),\, 
u_{\alpha 0}=\frac{1+i}{2}\sqrt{\frac{t}{m\hbar}}
\left(p_{\alpha}-\frac{m x}{t}\right).
\eeq
In an atomic fountain, trapped atoms are launched vertically with the help
of the optical molasses technique. 
PIAB eigenstates are a good approximation to 
the actual state of a box subjected to a linear potential, whenever 
the box is small and the particle light.
Indeed, within first order perturbation theory the coefficients $C_{nk}$
responsible for the corrections on the $n$-th eigenstate 
($\vert\phi_{n}^{(1)}\ra=\vert\phi_{n}^{(0)}\ra+\sum_{k\neq n}C_{nk}\vert\phi_{k}^{(0)}\ra$) 
due to the linear potential are given by
\beq
C_{nk}=\frac{8mfL^{3}}{\hbar^{2}\pi^{4}}\frac{nk[(-1)^{n+k}-1]}{k^{2}-n^{2}}.
\eeq

If the perturbation of the initial state by the linear potential is significant, 
it can always be written as a linear combination of the 
eigenstates of the free Hamiltonian within the box, namely,  
$\{\phi_{n}(x)=\sqrt{2/L}\sin(n\pi x/L) \vert n\in{\bf N}\}$. 
We must therefore look for the evolution of a PIAB state released 
at time $t=0$ and launched with a given momentum $q$.  
Explicitly, 
it is described by (\ref{bgrav}), with the redefinition
\beq
p_{\alpha}\equiv q+\frac{\alpha\hbar n\pi}{L}-\frac{f t}{2}.
\eeq
Equation (\ref{bgrav}) can model too the output coupling mechanism 
of an atom laser.
In such devices, an atom is initially in the lasing mode, confined by 
the cavity mirrors which  can be simulated by two infinite wall (reducing 
the problem to that of a PIAB). Through a Raman transition, this atom evolve 
to a nontrapped state, which amounts to shutting off the walls. 
During the transition and due to the emission of a photon the atom is kicked 
with a given momentum, namely, $q$ \cite{HDKWHRP99}.

\begin{figure}
\begin{center}
\includegraphics[height=6cm,angle=0]{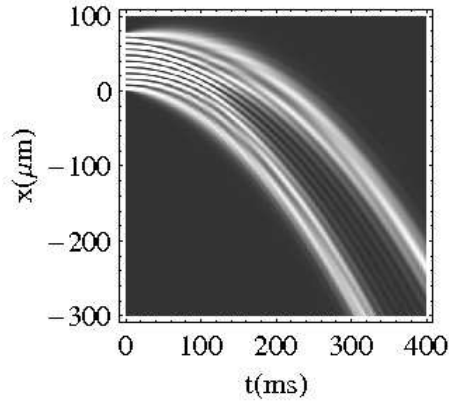}
\end{center}
\caption{\label{fountain10} 
Time evolution of the density profile for an eigenstate of the hard-wall trap, 
with $n=10$, $q=0$ cm/s, $L=80$ $\mu$m and $f/m=0.49$ cm/s$^{2}$.
Bifurcation for $t>t_{n=10}\simeq 137$ ms takes place in analogy with free case. 
The expectation value $\la\hat{x}(t)\ra$ reproduces the classical trajectory.}
\end{figure}
%

Figure \ref{fountain10} shows the time evolution of the density profile for 
the tenth eigenstate of the initial hard-wall trap. 
A most relevant feature is that at variance with the well-known harmonic trap case, 
the profile bifurcates into two main branches after the semiclassical time 
$t_{n}=m L/(2 p_{n})=mL^{2}/(2n\pi\hbar)$.
This behaviour has been recently reported for the free particle 
case by the authors \cite{DM05b} and holds for any excited state 
for both even and odd quantum number $n>1$.
The cloud falls under acceleration following 
$\la\psi\vert\hat{x}(t)\vert\psi\ra$ the classical trajectory, 
even though for any $t>t_{n}$ the probability is vanishing along it.


%
\section{Tonks gas dynamics in a linear potential}\label{tg}

Under strong radial confinement and at low densities and temperatures 
an ultracold atomic gas becomes effectively one dimensional.
In particular, whenever the thermal and zero-point energies are 
lower than the transversal excitation quantum, the Tonks-Girardeau (TG)
regime is reached. 
The effective interactions are then so strong that the gas behaves 
like a system of hard-core impenetrable bosons. 
Such regime has been obtained in several experiments \cite{exp} and 
a full quantum description is 
possible thanks to the Fermi-Bose mapping theorem \cite{Girardeau60}.
According to it, the many-body wavefunction of a TG gas can be obtained 
from the one of a zero-spin free fermionic gas, by applying the ``antisymmetric unit function''
$\mathcal{A}=\prod_{1\leq j<k\leq N}sgn(x_{k}-x_{j})$, as $\psi_{B}(x_{1},\dots,x_{N};t)=
\mathcal{A}(x_{1},\dots,x_{N})\psi_{F}(x_{1},\dots,x_{N};t).$
Indeed one can state $\vert\psi_{B}(x_{1},\dots,x_{N};t)\vert^{2}
=\vert\psi_{F}(x_{1},\dots,x_{N};t)\vert^{2}$, for the mapping is involutive.
The dual system to the TG gas, 
being the ideal Fermi system, is built up as a Slater determinant 
\beq
\psi_{F}(x_{1},\dots,x_{N};t)=\frac{1}{\sqrt{N!}}det_{n,k=1}^{N}\phi_{n}(x_{k};t).
\eeq
Moreover, thanks to the fact that under unitary time evolution, orthonormality between 
the initial eigenstates of the trap is preserved, it follows that any local 
correlation function can be obtained in close-form. For example, 
the calculation of the density profile is greatly simplified to
\beq
\label{dp}
\varrho(x,t)=N\!\!\int\vert\psi_{B}(x,x_{2},\dots,x_{N};t)\vert^{2}dx_{2}
\cdots dx_{N}
=\sum_{n=1}^{N}\vert\phi_{n}(x,t)\vert^{2}.
\eeq
The free expansion dynamics of a TG gas from a hard-wall trap was recently considered in \cite{DM05b}, 
where a dynamics much more complicated than for the harmonic confinement was observed.
Taking advantage of the results obtained in the previous section we next 
generalize the free expansion results and deal with the following experiment.
Consider a TG gas initially confined in a hard wall trap.
At zero time it receives a momentum kick, momentum at which the shutter is off 
and the linear potential rumped up.
The density profile is found combining (\ref{bgrav}) and (\ref{dp}).
\begin{figure}
\begin{center}
\includegraphics[height=6cm,angle=0]{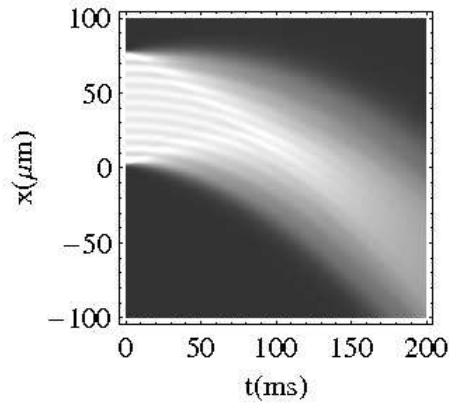}
\end{center}
\caption{\label{tonks10} 
Time evolution of the density profile for Tonks-Girardeau gas , 
with $N=10$, $q=0$ cm/s, $L=80$ $\mu$m and $f/m=0.49$ cm/s$^{2}$.
The interference pattern is washed out for $t>t_{N=10}\simeq 137$ ms.}
\end{figure}

Figure \ref{tonks10} shows the density profile of the expanding cloud 
in space and time falling under the action of a linear potential.
For short times $t<t_{N}$ there is a well-defined pattern where 
the number of maxima equals that of particles. 
After a transient regime where the visibility of the peaks varies 
in a non-trivial way, it is gradually lost for $t>t_{N}$. 
In the later regime, the ballistic expansion is established in agreement with 
the force-free case \cite{DM05b}.

%
%



\section{Conclusions}

Diffraction in time has been generalized in the presence of a 
linear potential.
Moreover, the dynamics of particles released 
from a hard-wall trap under a constant force has been studied, 
showing a bifurcation after a time $t_{n}=mL^{2}/(2n\pi\hbar)$, 
in analogy with the free case \cite{DM05b}. 
This is a novel feature exclusively
associated with  hard-wall traps. 
Note that for the case of harmonic confinement, 
the evolution follows a scaling law of coordinates, 
lacking therefore any transient effect \cite{PZ98}.
In addition, the expansion of a Tonks-Girardeau gas has been 
exactly solved using the Bose-Fermi map. The cloud is shown to exhibit an interference 
pattern which is lost for $t>t_{N}$.
Finally, we point out that any time dependence on the linear potential 
can be easily included using the general map obtained in \cite{Song03}.
In such a way, one can account for a smooth ramping up of the linear potential. 

\ack{
The papers has been benefited from discussions with A. Ruschhaupt and M. Rodriguez.
This work has been supported by Ministerio de Educaci\'on y Ciencia
(BFM2003-01003), and UPV-EHU (00039.310-15968/2004).
A.C. acknowledges financial support by the Basque Government (BFI04.479). 
}




\section*{References}
\begin{thebibliography}{10}
\expandafter\ifx\csname natexlab\endcsname\relax\def\natexlab#1{#1}\fi
\expandafter\ifx\csname bibnamefont\endcsname\relax
  \def\bibnamefont#1{#1}\fi
\expandafter\ifx\csname bibfnamefont\endcsname\relax
  \def\bibfnamefont#1{#1}\fi
\expandafter\ifx\csname citenamefont\endcsname\relax
  \def\citenamefont#1{#1}\fi
\expandafter\ifx\csname url\endcsname\relax
  \def\url#1{\texttt{#1}}\fi
\expandafter\ifx\csname urlprefix\endcsname\relax\def\urlprefix{URL }\fi
\providecommand{\bibinfo}[2]{#2}
\providecommand{\eprint}[2][]{\url{#2}}

\bibitem{Moshinsky52} Moshinsky M 1952 {\it Phys. Rev.} {\bf 88} 625

\bibitem{Moshinsky76} Moshinsky M 1976 {\it Am. Jour. Phys.} {\bf 44} 1037

\bibitem{DM05} del Campo A and Muga J G 2005 {\it J. Phys. A} {\bf 38} 9803 

\bibitem{Kleber94} Kleber M 1994 {\it Phys. Rep.} {\bf 236} 331 

\bibitem{GR97} Garc\'\i a-Calder\'on  G 
and Rubio A 1997 {\it Phys. Rev}. A {\bf 55} 3361

\bibitem{GV01} Garc\'\i a-Calder\'on  G 
and  Villavicencio J 2001 {\it Phys. Rev.} A {\bf 64} 012107

\bibitem{SSDD95} Steane A, Szriftgiser P, Desbiolles P and Dalibard J 1995
{\it Phys. Rev. Lett.} {\bf 74} 4972 

\bibitem{ASDS96} Arndt A, Szriftgiser P, Dalibard J and Steane A M 1996 
{\it Phys. Rev.} A {\bf 53} 3369

\bibitem{Salomon} Clairon A, Laurent P, Nadir A, Drewsen M, Grison D, Lounis B 
and Salomon C 1992 {\it EFTF: Proceedings of 6th European
Frequency and Time Forum} 

\bibitem{FKS02} Folman R, Kr{\"u}ger P and Schmiedmayer J 2002 
{\it Adv. Atom. Mol. Opt. Phys.} {\bf 48} 263 

\bibitem{MS99} Metcalf H J and van der Straten P 1999 
{\it Laser Coooling and Trapping} (New York: Springer) 

\bibitem{Kramer03} Kramer T, Bracher C and Kleber M 2002 {\it J. Phys. A} {\bf 35} 8361; 
Kramer T 2003 {\it Matter waves from localized sources in homogeneous force fields} 
http://tumb1.biblio.tu-muenchen.de/publ/diss/ph/2003/kramer.pdf
and reference therein



\bibitem{MSHCR05} Meyrath T P, Schreck F, Hanssen J L, Chuu C-S. and Raizen M G 2005
{\it Phys. Rev.} A {\bf R71} 041604 

\bibitem{GK76} Gerasimov A S and Kazarnovskii M V 1976 {\it Sov. Phys. JETP} 
{\bf 44} 892

\bibitem{Godoy02} Godoy S 2002  {\it Phys. Rev.} A {\bf 65} 042111

\bibitem{Gaudin71} Gaudin M 1971 {\it Phys. Rev.} A {\bf 4} 386

\bibitem{Cazalilla02} Cazalilla M A 2002 {\it Europhys. Lett.} {\bf 59} 793; 
Cazalilla M A 2004 {\it J. Phys.} B {\bf 37} S1

\bibitem{BGOL05} Batchelor M T, Guan X W, Oelkers N and Lee C 2005 
{\it J. Phys.} A {\bf 38} 7787

\bibitem{DM05b} del Campo A and Muga J G 2005 cond-mat/0511747

\bibitem{Mark} Raizen M G 2005 {\it Personal communications}

\bibitem{CMPL05} Colombe Y, Mercier B, Perrin H and Lorent V 2005 
{\it Phys. Rev.} A {\bf 72} 061601

\bibitem{Faddeyeva} Faddeyeva V N and Terentev N M 1961 
{\it Mathematical Tables: Tables of the values of 
the function $w(z)$ for complex argument} (New York: Pergamon)  
\bibitem{Abram} Abramowitz A and Stegun I A 1965
{\it Handbook of Mathematical Functions} (New York: Dover) 

\bibitem{Moshinsky51} Moshinsky M 1951 {\it Phys. Rev.} A {\bf 84} 525

\bibitem{GS98} Grosche C and Steiner F 1998 {\it Handbook of Feynman path integrals}, 
{\bf 145} {\it Springer Tracts in Modern Physics} (Berlin: Springer)




\bibitem{Wigner32} Wigner E P 1932 {\it Phys. Rev.} {\bf 40} 749

\bibitem{wigexp} Smithey D T {\it et al} 1993 {\it Phys. Rev. Lett.} {\bf 70} 1244;
Leibfried D {\it et al} 1996 {\it Phys. Rev. Lett.} {\bf 77} 4281;
Breitenbach G, Schiller S and Mlynek J 1997 {\it Nature} {\bf 387} 471;
Kurtsiefer Ch, Pfau T and Mlynek J 1997 {\it Nature} {\bf 386} 150;
Lvovsky A I {\it et al} 2001 {\it Phys. Rev. Lett.} {\bf 87} 050402;
Lougovski P {\it et al} 2003 {\it Phys. Rev. Lett.} {\bf 91} 010401

\bibitem{MMS99} M{\'{a}}nko V, Moshinsky M and Sharma A 1999
{\it Phys. Rev.} A {\bf 59} 1809 




\bibitem{HDKWHRP99} Hagley E W {\it et al} 1999 {\it Science} {\bf 283} 1706

\bibitem{exp} Paredes B {\it et al} 2004 {\it Nature (London)} {\bf 429} 227; 
Kinoshita  T,  Wenger T and  Weiss D 2004 {\it Science} {\bf 305} 1125

\bibitem{Girardeau60} Girardeau M 1960 {\it J. Math. Phys.} {\bf 1} 516

\bibitem{BM96} Brouard S and Muga J G 1996 {\it Phys. Rev.} A {\bf 54} 3055 

\bibitem{PZ98} Perelomov A M and Zel'dovich Y B 1998 {\it Quantum mechanics: selected topics}, 
(Singapore:World Scientific)

\bibitem{Song03} Song D Y 2003 {\it Europhys. Lett.} {\bf 65} 622

\end{thebibliography}
\end{document}